\documentclass[aps,prd,twocolumn,superscriptaddress,showpacs,preprintnumbers,amsmath,amssymb,floatfix]{revtex4}
\usepackage{epsfig}
\usepackage{graphics}
\input epsf.tex
\def\DESepsf(#1 width #2){\epsfxsize=#2 \epsfbox{#1}}

\newcommand{\be}{\begin{equation}}
\newcommand{\ee}{\end{equation}}
\newcommand{\bea}{\begin{eqnarray}}
\newcommand{\beas}{\begin{eqnarray*}}
\newcommand{\eea}{\end{eqnarray}}
\newcommand{\eeas}{\end{eqnarray*}} 
\newcommand{\ba}{\begin{array}}
\newcommand{\ea}{\end{array}}

\begin{document}

\title{Some ways of combining optimum interval upper limits}
\author{S. Yellin} \email{yellin@slac.stanford.edu}
\affiliation{Department of Physics, Stanford University,
Stanford, CA 94305, USA}
\date{\today}
\begin{abstract}

When backgrounds are not well enough controlled to measure the
value of some physical constant, one may still obtain an upper limit
on the constant.  A single experiment may have several detectors,
each of which can alone be used to derive an upper limit from the set
of detected events, and the experiment can be run during multiple periods.
There can also be more than one experiment which produces an upper limit.
Six methods are discussed for producing a combined upper limit.  These
methods all assume use of either the optimum interval method or the maximum
gap method for finding an upper limit in the face of poorly known background.

\end{abstract}

\maketitle

\section{Introduction}
In a measurement based on detection of events there can be ``signal'' events
from a physical process of interest characterized by the size of
some physical constant, and there can be ``background'' events
which individually cannot be distinguished from signal events,
but which come from other physical processes.
Although the Feldman-Cousins~\cite{FC}
method can be used to set limits when the uncertainty on the background
contribution is much smaller than the contribution from the signal,
that condition is not
necessarily met.  Instead the optimum interval method~\cite{yellin}
(or the simpler, but weaker maximum gap method) can be used to produce
an upper limit.  The event distribution in some measured quantity is used
to obtain
an upper limit on the size of the physical constant that could produce
the signal.  If the event distribution differs
from what would be expected from the signal alone, the maximum
gap and optimum interval methods can get stronger limits than would be
obtained by simply counting events.  They use an interval in the
range of the measured quantity that gets the strongest (lowest) upper
limit, while taking the proper statistical penalty for the freedom to
choose an interval which is especially free of background.  The methods
involve first transforming the measured quantity of each event into
the cumulative probability of its value -- the probability of
a random event from the assumed signal having a lower value of the measured
quantity
than the observed event has.  In the absence of background, the probability
distribution of this transformed quantity is uniform between zero and unity.
The optimum interval and maximum gap methods take advantage of any
non-uniformity in the background to find an upper limit on the signal
by using, in a statistically proper manner, an interval with especially
few events within the range of this transformed quantity.  An example
considered in various parts of this paper is the
search for weakly interacting massive particles (WIMPs) which pass through
the Earth while orbiting within the Milky Way.  The signal size corresponds
to the value of the WIMP cross section on nucleons.
The measured quantity for each event is the energy each WIMP candidate
deposits in a detector.  The shape of the expected WIMP energy spectrum can
be computed for each assumed WIMP mass, this shape can be integrated
to get the cumulative probability distribution expected for WIMPs of each
mass, and the
optimum interval method can then be used to set a cross section upper limit as
a function of WIMP mass.

This paper discusses various ways of using the optimum interval method
when combining measurements to produce a
combined upper limit.  Most of the discussion will be for
exactly two measurements, although all methods can in principle, and sometimes
in practice, be generalized to an arbitrary number.
The method chosen should be selected based on the circumstances of
the measurements, but as much as feasible be selected ``blindly'', without
testing which method gets the strongest limit -- it
would bias the result low to try several methods and choose the one that
produces the strongest limit.  We require each method to be statistically
valid -- for example a 90\% confidence level upper limit must be obtained
by a method whose probability of mistakenly excluding the truth is no more
than 10\%.  Monte Carlo simulations with software~\cite{software} implementing
all discussed methods have confirmed this property.

We would like the
methods to have two virtues: A) get a stronger limit than either measurement
alone if the data justify it, and B) be robust against one of the two having
worse backgrounds than the other.  To get an idea of the extent to which each
method has these virtues, we will test their performance under two scenarios.
For both scenarios assume that two measurements are made by two experiments
with the same measured quantity range and sensitivity,
and that there is no signal.  Assume there is no background in
experiment 2.  Test ``A'' checks for virtue A by seeing how strong a
limit results if experiment
1 also has no background, and test ``B'' tests for virtue B by seeing
how much experiment 2's limit is
hurt if it's combined with an experiment 1 suffering from an extremely
large background over its entire range.  The first three of the
methods discussed perform well on test A, but very badly on test B.
The fourth performs well on test A, and better than the first three on B.
The last two are designed for use with high backgrounds in both experiments,
but their performance can also be investigated for the two low background 
tests.  They perform better than the methods appropriate for low backgrounds
on test B, but not test A.

Another test, called here ``test C'', uses CDMS low background
data for each of the six methods.
The CDMS experiment~\cite{CDMS}
used very low temperature germanium crystals in an attempt to detect both
the sound and the ionization from WIMPs scattering on nuclei.  We imagine the
CDMS exposure split into two equal halves of identical effective net
exposure and efficiencies, with the WIMP candidate
events arbitrarily allocated between the two halves.  Each method for
combining experiments is used to combine the two halves for all $2^4$ possible
ways of assigning events to the two halves.

Since A, B, and C don't properly test performance in the face of
high backgrounds, the last two methods are also challenged with
data from a CDMS high background experiment.

It is up to the user
to choose a method based on what is known in advance about the backgrounds.

\section{Methods for Low Backgrounds}
\subsection{Simple Merging}
We begin with a method that is especially easy to understand and implement.
If two or more measurements are almost
background free then each of their
limits could benefit from a larger exposure.
In that case if they're sufficiently alike a good way to combine them is
the ``simple merging'' method: treat them as one bigger experiment.
Combine the events from the measurements into one set of events, giving
a single event distribution as a function of the measured quantity used,
and compute the expected spectrum of the combined set of events.  Then use
the optimum interval method to compute a joint upper limit.
Simple merging has been used by the
CDMS~\cite{CDMS} collaboration to combine WIMP search results from its
germanium detectors, and has also been used for the same purpose by the
EDELWEISS~\cite{EDELWEISS} and CRESST~\cite{CRESST} collaborations.

Although simple merging can
be used to combine data from different types of detectors, extra
care must be taken
to do it well.  Consider, for example, the following scenario: One wishes to 
get an upper limit for WIMPs of mass 20 GeV/$c^2$ from combining an
experiment with Ge detectors and an experiment which uses Xe, and 
one of the Xe events has recoil energy of 40 keV.
Take the maximum velocity of WIMPs bound within the Galaxy to be $\sim 544$
km/s~\cite{HaloModel}, from which it follows that a WIMP of mass 20 GeV/$c^2$
cannot produce an elastic scatter in Xe with more recoil energy than about
33 keV.  The 40 keV event must therefore be from background.  The optimum
interval method applied to Xe alone would correctly ignore this event
because its transformed quantity couldn't be in the optimum interval; there
is probability unity of a WIMP event depositing less energy than 40 keV, so
the event would be at the very end of the transformed range.  But such a WIMP
could produce an event with recoil energy as high as 48 keV in
Ge, so if recoil energies are merged, the 40 keV background event would
not be at the very end of the transformed range of the merged experiments,
and could make the upper limit unnecessarily conservative.  Such
a problem can be prevented by a variation of simple merging,
``cumulative probability merging''.  Instead of merging recoil energies, those
energies are first transformed into cumulative probabilities for
each experiment, and then the transformed quantities are merged.
Cumulative probability merging would put
the 40 keV Xe event at the end of the merged range, where it
would have no effect on the merged upper limit.

A nice feature of cumulative probability merging
is that the same upper limit would result if one or both experiments used
some quantity other than recoil energy.  If, to pick a strange example,
the Ge experimenters expressed their events in terms of recoil momentum,
not energy, while the Xe experimenters
described their events by recording in place of recoil energy the
minimum WIMP mass that could produce an elastic
scatter with the observed recoil energy, the transformed values for each
experiment would be the same probabilities as if both expressed their
data in terms of recoil energy.  In fact, experiments may be merged
which differ so greatly that they don't even measure the same physical
characteristic of the event.  All other methods described in this paper for
combining measurements automatically have the same ability.

For the conditions of test A, simple merging
results in one combined measurement with twice
the exposure, but still no events, so the upper limit is made stronger by
a factor of 1/2.  For the conditions of test B, however, experiment 1 badly
pollutes the otherwise clean experiment 2 data, resulting in a very much
weaker upper limit.  For test C, both simple merging and cumulative probability
merging of the two half CDMS
experiments are equivalent to the method used for the published CDMS
result~\cite{CDMS} from simple
merging of data from all well functioning detectors.

\subsection{Combining with Confidence Levels}
For most methods discussed in this paper, measurements are combined based
on upper limit confidence levels from them, rather than from using their
events directly as in simple merging.  
For a given value of the physical constant being sought, call ``$\mu$'' the
expectation value of the number of events from the
physical process.  Take ``$p$'' to be the confidence level by which the
measurement excludes a particular value of the physical constant.  If this
confidence level is found by the maximum gap or optimum interval method,
$p$ itself has a certain probability distribution under the assumption of
no background, and with correct assumptions about the physical process whose
strength is being measured.  In order to see what the probability
distribution is, imagine a huge number of independent similar
experiments, all having no background and all having the same true value
of the physical constant corresponding to $\mu$ expected number of events.
These could be generated by a Monte Carlo simulation.
For experiment $i$, the $p$ confidence level upper limit
$\mu_i(p)$ can be computed.  By the definition of ``confidence level'',
$p=0.9$ means that for 90\% of the experiments
$\mu < \mu_i(p)$, while there is only a 10\% probability that the
upper limit $\mu_i(p)$ is mistakenly found to be higher than $\mu$.  In
general, $p$ is the probability of $\mu < \mu_i(p)$.  In a region of $p$
where the probability distribution of $p$ is continuous and
non-zero, the probability that
$\mu < \mu_i(p+dp)$ is $p+dp$ and the probability that $\mu < \mu_i(p)$
is $p$, so there is probability $dp$ that $\mu$ is between $\mu_i(p)$ and
$\mu_i(p+dp)$.  There is probability $dp$ of the true value of the constant
being between the $p$ and $p+dp$ confidence level values,
so the probability distribution of $p$ has $\rm{d(Probability)}/dp = 1$
in regions where the probability density of $p$ is non-zero and continuous. 
The probability density is zero at values of $p$ for which none of the
experiments produces that confidence level, as when $p<0$ or $p$
is greater than its maximum possible value.  There is probability
$e^{-\mu}$ of there being zero events in the entire range of measurement,
in which case the optimum interval is the entire
range of the measurement, and the corresponding $p$ is at its maximum
possible value, $1-e^{-\mu}$.  Thus the probability density of $p$ is
zero for $p > 1-e^{-\mu}$, and the density has a delta function at
$p=1-e^{-\mu}$ whose integral is $e^{-\mu}$.
To summarize, if $p$ is correctly computed for the
optimum interval method (or for the maximum gap method), then in the absence
of background it has probability zero of being below zero or above
$1-e^{-\mu}$, has probability $e^{-\mu}$ of being exactly $1-e^{-\mu}$,
and is otherwise uniformly distributed with unit density for
$0 < p < 1-e^{-\mu}$.

Maximum gap or optimum interval results can be combined to
form some function $q(p_1,p_2,\mu_1,\mu_2)$ in such a way
that if the observed $q$ is too large for an assumed value of the physical
constant being investigated, the value
will be excluded.  A value for a physical constant is conservatively excluded
to the 90\% confidence level when it is so large that a random pair of
measurements similar to measurements 1 and 2, and with no unknown background
adding to the signal, would have a
90\% probability of $q$ being smaller than was observed.  The probability
of $q$ being smaller than was observed can be computed from the
probability distributions of $p_1$ and $p_2$.  Different
methods correspond to different choices for $q$ as a function of results
from the two measurements.  Choose measurement 1 to be the one with the
larger expected number of events: $\mu_1\ge\mu_2$.

\subsection{Minimum Probability}
The ``minimum probability'' method takes $q$ to be the
smaller of $p_1$ and $p_2$.
The probability of $q$ being smaller than $z$ is
\be
\begin{array}{cll}
P_{MinP}(z,\mu_1,\mu_2) &= 0 & z\le 0 \\
                   &= 2z-z^2 & 0\le z < 1-e^{-\mu_2} \\
                   &= 1 & z\ge 1-e^{-\mu_2} \\
\end{array}
\ee
$P_{MinP}$ jumps by $e^{-2\mu_2}$ at $z=1-e^{-\mu_2}$.

For test A the minimum probability upper limit of the combined experiment
improves the single experiment limit by a factor of 1/2.  But this method
completely fails test B; the combined experiment increases the upper
limit by a large factor.

\begin{figure}[ht]
        \centering
        \epsfig{file=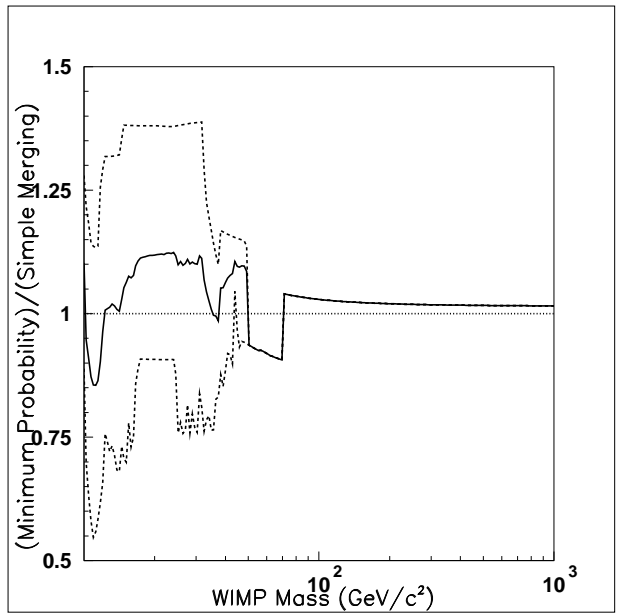,width=\linewidth}
        \caption{Ratio of the minimum probability upper limit to the simple
merging limit from combining two equal halves of CDMS~\cite{CDMS}.  The solid
curve is the ratio averaged over all ways of allocating events between the two
halves, and the dashed
curves are the maximum and minimum values of the ratio as the way of dividing
up the events between the two halves is varied.}
        \label{fig_vs5}
\end{figure}

The result of test C for the minimum probability upper limit is shown
in Fig.~\ref{fig_vs5}.  The figure shows the ratio between the minimum
probability upper limit curve and the simple merging upper limit curve
for combining two equal halves of the CDMS experiment~\cite{CDMS}.  When
the ratio is below the horizontal dotted line at 1.0, the minimum probability
upper limit is stronger than the already published simple merging one.

\subsection{Probability Product}
The ``probability product'' method takes
$q = p_1p_2$.  The probability that $q$ is smaller than $z$ is
$P_{ProdP}(z,\mu_1,\mu_2) =$

\be
\begin{array}{ll}
0 & z\le 0\\
z\,log\frac{(1-e^{-\mu_1})(1-e^{-\mu_2})}{ z}\\
\ \ \ + \frac{z(1-e^{-(\mu_1+\mu_2)})}{ (1-e^{-\mu_1})(1-e^{-\mu_2})}&
        0 \le z < z_{max} \\
1 & z\ge z_{max} \\
\end{array}
\ee
where $z_{max} = (1-e^{-\mu_1})(1-e^{-\mu_2})$.  At $z=z_{max}$
$P_{ProdP}$ jumps by $e^{-(\mu_1+\mu_2)}$. 

For test A the probability product upper limit of the combined experiments
improves the 90\% confidence level upper limit by a factor of 1/2.
This method, like the other methods discussed so far, completely fails test B.

\begin{figure}[tbp]
        \centering
        \epsfig{file=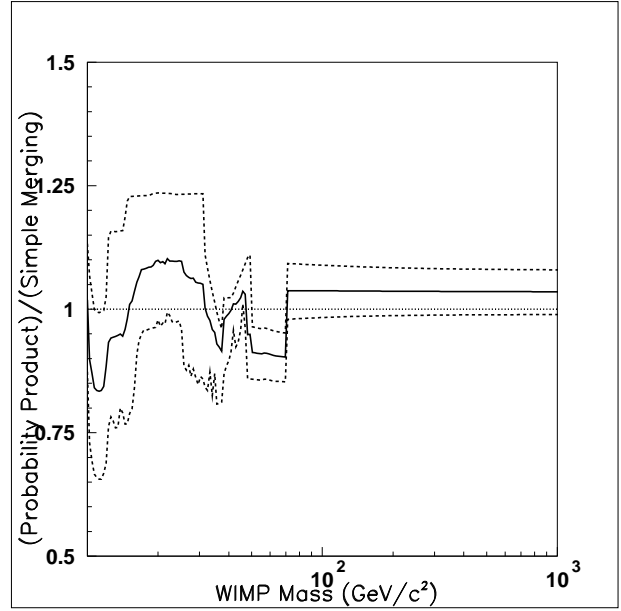,width=\linewidth}
        \caption{Ratio of the probability product upper limit to the simple
merging limit from combining two equal halves of CDMS~\cite{CDMS}.    The solid
curve is the ratio averaged over all ways of allocating events between the two
halves, and the dashed
curves are the maximum and minimum values of the ratio as the way of dividing
up the events between the two halves is varied.}
        \label{fig_vs6}
\end{figure}

The result of test C for the probability product upper limit is shown
in Fig.~\ref{fig_vs6}.

\subsection{Summed Maximum Gap}
We next discuss a method which should be effective when the backgrounds of
separate measurements are low, but which unlike the other methods discussed
so far is not too badly hurt if
the background of one is much worse than that of another.   Suppose
two measurements have their upper limits determined using the maximum gap
method.  A ``gap'' between neighboring events has its size characterized
by ``$s$'', the expectation value of the number of events from the signal.
The maximum gap, ``$x$'', is the maximum over all gaps of $s$.
A ``summed maximum gap'' for two
measurements takes $q=x_1+x_2$, where $x_1$ and $x_2$ are the maximum gaps
of the two measurements.  If the two measurements have their upper
limits determined by the optimum interval methods, $x_1$ and $x_2$ can instead
be taken as the maximum gaps that would give the same confidence levels
as the optimum interval method gave.  In terms of the $C_0$ function defined
in reference \cite{yellin},
the ``summed maximum gap'' method takes $q=x_1+x_2$, where $x_1$
and $x_2$ are $X_0(p_1,\mu_1)$ and $X_0(p_2,\mu_2)$, and $X_0(p,\mu)$ is
an inverse function to $C_0(x,\mu)$;
it's the value for which $C_0(X_0(p,\mu),\mu)=p$.

The probability that a random pair of background-free measurements
will give $q$ smaller than some value, $z$, given $\mu_1$ and $\mu_2$ is
$P_{SumG}=$

\be
\begin{array}{ll}
0  & z\le 0 \\
\int_0^z dx_1 \frac{\partial C_0(x_1,\mu_1)} {\partial x_1}
        C_0(z-x_1,\mu_2) & 0\le z \le \mu_2 \\
\int_{z-{\mu_2}}^z  \frac{\partial C_0(x_1,\mu_1)}
        {\partial x_1} C_0(z-x_1,\mu_2)\\
\ \ \  +C_0(z-\mu_2,\mu_1) &  \mu_2\le z \le \mu_1 \\
\int_{z-{\mu_2}}^{\mu_1} \frac{\partial C_0(x_1,\mu_1)}
         {\partial x_1} C_0(z-x_1,\mu_2)\\
\ \ \ +C_0(z-\mu_2,\mu_1)\\
\ \ \ + e^{-\mu1}C_0(z-\mu_1,\mu_2)  & \mu_1 \le z \le \mu_1+\mu_2 \\
1 & z > \mu_1+\mu_2\\
\end{array}
\ee

$P_{SumG}$ has a discontinuity at $z=\mu_1+\mu_2$; it jumps by $e^{-(\mu_1+\mu_2)}$.
Although the integrals can be evaluated in closed form, those forms haven't
yet been made simple enough to use.  But $P_{SumG}(z,\mu_1,\mu_2)$ can be
evaluated numerically, and possibly
tabulated.  The 90\% confidence level upper limit cross section is
the cross section for which $0.9 = P_{SumG}(x_1+x_2,\mu_1,\mu_2)$.

Test A for the summed effective maximum gap method  gives a factor of 1/2
improvement in the upper limit
compared to the limit for each experiment alone.  For test B, the 90\%
confidence level upper limit for $\mu$ is at $\mu=6.679$, the solution of
\be
0.9 = \int_0^\mu dx \frac{\partial C_0(x,\mu)} {\partial x}C_0(\mu-x,\mu).
\ee
The 90\% confidence level upper limit is therefore increased by a factor of
$6.679/2.303 = 2.90$ over its value for experiment 1 alone.

\begin{figure}[tbp]
        \centering
        \epsfig{file=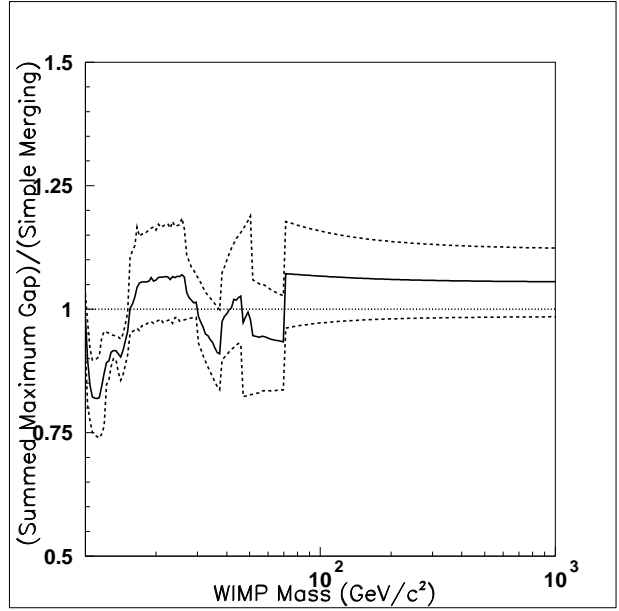,width=\linewidth}
        \caption{Ratio of the summed effective maximum gap
upper limit to the simple
merging limit from combining two equal halves of CDMS~\cite{CDMS}.  The solid
curve is the ratio averaged over all ways of allocating events between the two
halves, and the dashed
curves are the maximum and minimum values of the ratio as the way of dividing
up the events between the two halves is varied.}
        \label{fig_vs3}
\end{figure}

The result of test C for the summed effective maximum gap upper limit is shown
in Fig.~\ref{fig_vs3}.

\section{Methods for High Backgrounds}
We next discuss two methods intended for the case of large backgrounds.
For a particular set of physical assumptions, the background of
one measurement can be much worse than the background of the other one.
It may be that because the different
measurements have different backgrounds, one gets a stronger limit for
some sets of physical assumptions and another gets a stronger limit
for other sets of physical assumptions.  This situation is
likely if the different measurements are results from different
parts of the detection apparatus.  For example, the CDMS collaboration
normally combines results from many detectors using the simple merging
method~\cite{CDMS}.  But it has also lowered its analysis threshold and
modified its cuts to
provide improved sensitivity to lower WIMP masses
despite the high and uncertain backgrounds at low energies~\cite{Bunker}\cite{Moore}.
With a low threshold, detectors with especially low background could give
a stronger limit alone than they could when merged with detectors with
poorer background rejection.  And because the backgrounds were
differently distributed in the different detectors, which detector gave
the strongest limit could depend on the WIMP mass being tested.  The
limit would be biased low by simply choosing the lowest one for each WIMP
mass, because the lowest one could benefit from
a statistical fluctuation, rather than from superior
background rejection.  There must be a statistical penalty
paid for allowing the choice.  If the background is high, the statistical
penalty will be a small fraction of the upper limit.  High backgrounds may
require the high statistics version~\cite{yellin2} of the optimum interval
method.  If the differences
between backgrounds in different measurements are large enough,
the advantage from making a choice will be greater than the statistical
penalty that must be paid.

\subsection{Serialization}
\begin{figure}[tbp]
        \centering
        \epsfig{file=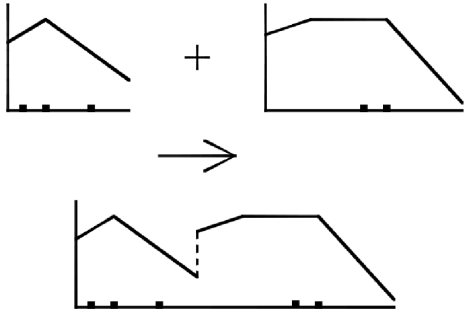,width=\linewidth}
        \caption{Diagram illustrating the serialization method.  The top
two plots portray two different measurements, each with its own
events displayed as small squares on the bottom of their plot, and each
with its own expected densities for detecting events from the process being
investigated, shown as continuous functions.  The events and expected
density distributions
are then put end to end, as shown in the bottom plot.}
        \label{fig_seq}
\end{figure}

A simple way to automatically choose the measurement with
best background rejection is
``serialization'': concatenate the measured quantity
ranges and apply the optimum interval method to the resulting total range.
The concatenation is illustrated in Fig. \ref{fig_seq}.

The optimum interval will include the part of the measurement range
which gives the strongest limit, and will therefore tend to use the
measurement with the strongest limit.  Because the optimum interval
method automatically takes the correct statistical penalty for its
freedom to choose the interval, serialization automatically takes
the correct statistical penalty for its freedom to choose as its optimum
interval one mainly or entirely in the part of the concatenated range for
the measurement which seems to be most free of background.  The result of
this method may depend on the order in which the measurements are
concatenated, but no additional statistical penalty need be taken
for the freedom to choose the order provided the choice
is made ``blindly'', without knowledge of which order gives the strongest
limit.  None of the other methods discussed in this paper depends on the
way measurements are ordered.

  For both
tests A and B, $ \mu_1=\mu_2\equiv\mu$, and the
optimum interval 90\% confidence level upper limit for experiment 2 alone
corresponds to a signal size for which $\mu=-log(0.1) = 2.30$.
Serialization of the two data sets under the conditions
of test A results in one combined measurement with twice the exposure,
but no events, resulting in an upper limit improved by a factor of 1/2.
For test B the 90\% confidence limit on $\mu$ becomes the $\mu$ for which
$\bar C_{Max}(0.9,2\mu) = C_0(\mu,2\mu)$ after serialization (notation
of Ref.~\cite{yellin}), resulting in 
a factor of $4.78/2.30 = 2.1$ increase (weakening) of the upper limit from
experiment 2 alone.  So while the background of experiment 1 hurts, it
doesn't hurt as much as for simple merging.  But if experiment 1 is the first
in the series and scenario A is modified to have many background events at
the bottom of the energy range of experiment 2, then instead of a factor of
1/2 strengthening of the upper limit, serialization will suffer approximately the
same factor of 2.1 weakening of the limit in the modified test A as in
test B.  Even modifying scenario A by adding just one background event at the
bottom of the range of experiment 2 will result in the combination producing
its upper limit where $1-e^{-2\mu}(1+2\mu) = \bar C_{Max}(0.9,2\mu)$, or
$\mu=2.167$, a factor of 0.94, instead of 0.50, of the experiment 2 limit
alone.  None of the other methods discussed in this paper are
significantly weakened by background events very near threshold.

\begin{figure}[tbp]
        \centering
        \epsfig{file=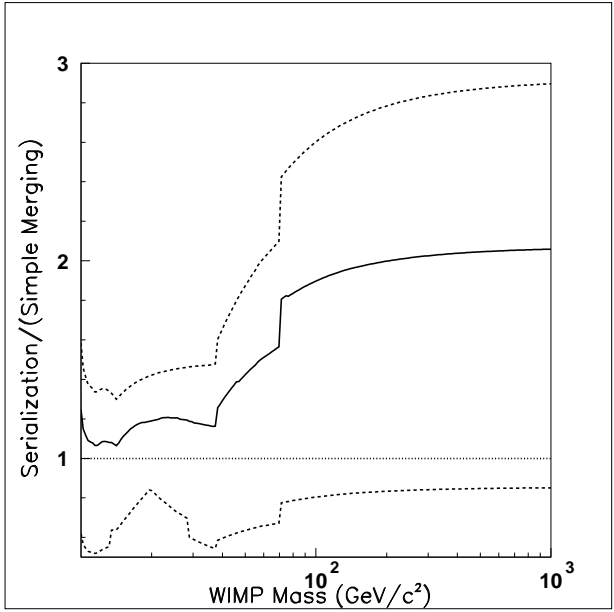,width=\linewidth}
        \caption{Ratio of the serialization upper limit to the simple
merging limit from combining two equal halves of CDMS~\cite{CDMS}.  The solid
curve is the ratio averaged over all ways of allocating events between the two
halves, and the dashed
curves are the maximum and minimum values of the ratio as the way of dividing
up the events between the two halves is varied.}
        \label{fig_vs2}
\end{figure}

The result of test C for the serialization upper limit is shown in
Fig.~\ref{fig_vs2}.  The previously discussed methods all give on the average
about the same upper limit as simple merging, depending on WIMP mass, but the
serialization upper limit for test C is on the average weaker than the simple
merging one, and for some ways of distributing the events is much weaker.

\begin{figure}[tbp]
        \centering
        \epsfig{file=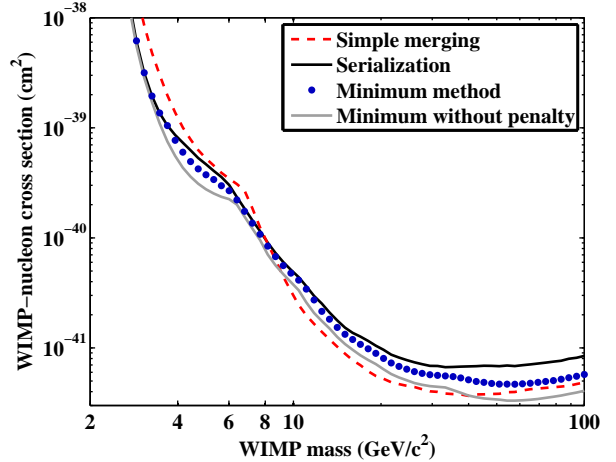,width=\linewidth}
        \caption{Comparison~\cite{Bunker2} of high background methods applied
to CDMS low threshold Ge data~\cite{Bunker}.  The dashed curve shows
the 90\% CL upper limit
from simple merging, the dark solid curve shows the result from the serialization
method, the dotted curve is for the minimum limit, and the light solid curve
shows what the minimum limit would be if there had been no penalty for the
freedom to choose the detector with the minimum limit.}
        \label{fig_Bunker}
\end{figure}

Tests A, B, or C are inappropriate for the recommended use of the
serialization method.  They all
have at least one experiment or detector with low background,
while the serialization method is intended for use with high background.
To remedy this deficiency we demonstrate the performance of the serialization
method with high background data, a CDMS low threshold analysis~\cite{Bunker}.
Figure \ref{fig_Bunker}~\cite{Bunker2} allows comparison between
simple merging and serialization for a 90\% confidence
level upper limit reanalysis of CDMS data taken at a shallow site.
Although the published results~\cite{Bunker} combined germanium and silicon
detectors, the silicon detectors had too much background and too little mass
for their inclusion to improve the combined limit.  Only
germanium detectors were used for Fig.~\ref{fig_Bunker}, which compares
merging with
concatenation of six measurements.  This figure shows
that at the low WIMP masses for which the reanalysis was performed and for
which the high background is important, the serialization method gives a
stronger limit than does simple merging.  At high masses, the optimum
interval occurs where backgrounds are low, and simple merging is a more
sensitive way of combining detectors.

\subsection{Minimum Limit}
The statistical penalty required by the serialization method must also
include a penalty for the possibility that the optimum interval may cross
the boundary between the ranges for two measurements.  The reason for using
intervals, rather than some other non-contiguous point set, is because
backgrounds tend to be larger in some parts of the experimental range than
others.  For example, the background may be especially large at the low
end of the experimental range, where detector signals are especially small.
Concatenating experimental ranges may well put a small background region
adjacent to a large background one, in which case the optimum interval
method would not choose an interval which extends much past the point at
which the two ranges join, even though it must take a penalty for the
possibility of doing so.  Under such circumstances, a somewhat stronger
limit can be obtained by instead directly choosing the measurement
which gives the strongest limit at a given confidence level, while taking
the correct statistical
penalty for the freedom to make such a choice: the ``minimum limit''
method.

Assume the confidence level by which an assumed physical constant
is excluded as too high decreases, or at least doesn't increase,
when the assumed value decreases.  This monotonic decrease of confidence
level with value of the constant has small exceptions for the optimum
interval method, but let's neglect such effects.  When measurements
with higher upper limits than the strongest one have the upper limit
value of the physical constant lowered until it's the same
as the strongest limit, the confidence levels
by which they're rejected will be lowered.  Thus finding the minimum limit
for the same confidence level is equivalent to finding the maximum confidence
level for the given upper limit.  ``Minimum limit'' is equivalent to
``maximum probability''.  For two measurements, $q$ is the larger of $p_1$
and $p_2$.   More generally, $n$ results are combined by taking $q$
to be the largest of all the probabilities, $p_i$.  If the highest $p_i$
for a given cross section is smaller than $z$, then all the $p_i$ are
smaller than $z$.  The probability of all $p_i$ being smaller than $z$
is equal to the product of the probabilities of each having $p_i<z$.
I.e., it is 0 if $z<0$, 1 if $z\ge 1$, and is otherwise
$z^k$, where $k$ is the number of measurements with $1-e^{-\mu_i} > z$.
The resulting function has discontinuities wherever $k$ changes.  Here's
an intuitive justification for lowering $k$ when it results in a measurement
with $\mu_i < -log(1-z)$, corresponding to $1-e^{-\mu_i} > z$.  Suppose, for
example, the maximum probability for a particular cross section is 96\%,
and for that cross section measurement $i$ has $\mu_i =3$.  Then measurement
$i$ cannot get a 96\% upper limit no matter what events are in it, because the
exclusion cross section from measurement $i$ cannot be greater than what one
would get from zero events, for which the confidence level would be
$1-e^{-3} = 95\% < 96\%$.  In that case, measurement $i$ shouldn't
be included in the penalty that must be paid for the multiple
chances of accidentally getting a maximum probability as high as 96\%.

The algorithm for finding
the, say, 90\% confidence level upper limit starts with finding the upper limit
for each measurement to the $z$ confidence level, where $z^k=0.9$, and
initially $k=n$.  If for that value of $z$
the number of measurements with $1-e^{-\mu_i} > z$ is $k$, conclude the
algorithm by taking the minimum upper limit as the overall 90\% one.
Otherwise repeat with $k\leftarrow k'= k-1$  until the number is $\ge k'$.
If the number is $>k'$ then there is no $z$ for which the confidence level for
rejection is exactly 90\%.  Conservatively choose the lowest confidence level
above 90\% for which there is a $z$ which rejects it.  I.e., find the lowest
$z$ for which there are $k'+1$ measurements with $1-e^{-\mu_i} \ge z$.

For both tests A and B the 90\% CL upper limit of $\mu=2.30$ from experiment
2 alone becomes $\mu=2.97$ from the combined measurement.

\begin{figure}[tbp]
        \centering
        \epsfig{file=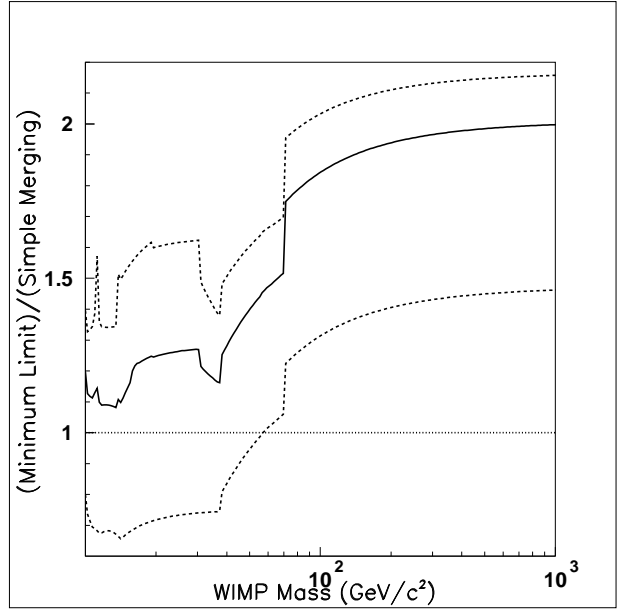,width=\linewidth}
        \caption{Ratio of the minimum limit upper limit to the simple merging
limit from combining two nearly equal halves of CDMS~\cite{CDMS}.  The solid
curve is the ratio averaged over all ways of allocating events between the two
halves, and the dashed
curves are the maximum and minimum values of the ratio as the way of dividing
up the events between the two halves is varied.}
        \label{fig_vs4}
\end{figure}

The result of test C for the minimum limit upper limit is shown
in Fig.~\ref{fig_vs4}.  The decision on whether a penalty is needed for the
existence of a second experiment can be ambiguous if the two halves are
exactly equal in exposure, so the two halves of the CDMS exposure were
made very slightly unequal.  As for the serialization method, the minimum limit
test C result is on the average weaker than the simple merging one.  Although
the minimum limit result can be stronger than the one from simple merging,
unlike the other methods discussed in this paper it cannot be stronger than
the strongest of the limits from the individual measurements.

Like the serialization method, the minimum limit method has been
tested\cite{Bunker2} on CDMS low threshold data~\cite{Bunker} taken
at a shallow site.
Figure \ref{fig_Bunker} shows that for this experiment the minimum limit is
somewhat stronger than the serialization one, probably because it need not
take a penalty for the possibility of the optimum interval overlapping two
detectors.  As for the serialization method, the minimum limit gives the
strongest limit at low masses, where backgrounds are high.  The lighter solid
curve in the figure shows by its difference with the dotted curve how much
of a penalty had to be taken for the freedom to choose the detector which gave
the minimum limit.  If CDMS had been able to tell in a blind
manner, without seeing the events actually used for setting a limit,
which detector to use for each WIMP mass, the light solid curve would have
been the upper limit.

\section{Conclusions}

Six methods have been discussed for obtaining a combined upper limit for
multiple experiments, or parts of experiments, which may individually
be analyzed using the optimum interval method.  The combination methods
also apply if the individual measurements are analyzed using the maximum
gap method.  None of the methods require that the different experments
use similar detectors or procedures, so long as they both can
set an upper limit on the same physical constant.
Some combination methods are especially useful if the individual
measurements have low enough background to be limited in sensitivity by the
amount of running time each has had.  Others are most useful if the individual
measurements have high and different backgrounds, and one wishes to use only
the most sensitive one while paying the appropriate statistical penalty for
the possibility that the apparently best one may seem more sensitive simply
because of a statistical fluctuation.  The various combination methods
were briefly examined for their ability to achieve sensitivity when
two experiments have very low background, and they were examined for their
sensitivity when only one experiment has
low background, while the other has very high background.  Qualitative
results are shown in Table \ref{summary_tab}.  This table applies mainly to
methods useful for combining measurements which have low backgrounds.
The methods appropriate for high backgrounds were tested on high background
experimental data.

Software is publicly available~\cite{software} for applying the methods
discussed in this paper.

\acknowledgments
{Bernard Sadoulet suggested thinking about ways of combining upper limits
from the CDMS and EDELWEISS collaborations.  This resulting paper has
been improved by advice from others, including Richard Schnee
and Philippe Di Stefano.}

\begin{table}[ht]
\caption{Sensitivity of the combination methods for two sample experiments,
each with either low or high backgrounds.}
\centering
\begin{tabular}{|c||c|c|}
\hline
Method & Both Low& One Low,\\
& & One High\\
\hline
Simple Merging& Good & Bad\\
Minimum Probability& Good & Bad\\
Probability Product& Good & Bad\\
Summed Maximum Gap& Good & Fair\\
Serialization & Fair & Fair\\
Minimum Limit& Fair & Fair\\
\hline
\end{tabular}
\label{summary_tab}
\end{table}

\end{document}